\documentstyle[psfig]{mn}

\ifnfsstwo

\fi

\ifnfssone
  \newmathalphabet{\mathit}
    \addtoversion{normal}{\mathit}{cmr}{m}{it}
    \addtoversion{bold}{\mathit}{cmr}{bx}{it}

\fi

\ifoldfss

\fi

\loadboldmathitalic
\loadboldgreek

\title{$UBV(RI)_{\rm C}$ photometry and spectroscopy\\
of the young open cluster Haffner~18
\thanks{Based on observations collected with the telescopes 
of the  South African Astronomical Observatory (Sutherland, RSA) and
ESO (La Silla, Chile).}}

\author[Ulisse Munari, Giovanni Carraro and Roberto Barbon]
       {Ulisse Munari$^{1}$ , Giovanni Carraro$^{2,3}$ and Roberto Barbon$^{4}$ \\
        $^{1}$) Osservatorio Astronomico di Padova, sede di Asiago,
 I--36012 Asiago (VI), Italy (\/{\tt munari\char64astras.pd.astro.it})\\
        $^{2}$) Dipartimento di Astronomia, Universita' di Padova,
 vicolo dell'Osservatorio 5, I--35122, Padova ,Italy\\
        $^{3}$) SISSA/ISAS, via Beirut 2, I--34013, Trieste, Italy
(\/{\tt carraro\char64sissa.it)}\\
        $^{4}$) Osservatorio Astrofisico, I--36012 Asiago (VI), Italy
(\/ {\tt barbon\char64astras.pd.astro.it
})}
\date {Accepted.......
      Received.......;
      in original form ......}
\pubyear{1995}

\begin{document}

\maketitle
\title{The young open cluster Haffner~18}

\begin{abstract}

{\sl UBV(RI)$_{\rm C}$} CCD photometry of a $2.^{\prime}1 \times
3.^{\prime}3$ field centered on the young open cluster Haffner~18 is
presented and discussed. Spectroscopic classification of seven stars is also
provided. 44 clusters members are identified, the earliest type being O6.
The distance to the cluster is found to be 6.3 kpc, corresponding to a
galactocentric distance of 12.7 kpc (for a Sun galactocentric distance of
8.5 kpc). Excellent fit to the observed main sequence is achieved by a solar
composition isochrone of 2$\times 10^6$ years reddened by E$_{B-V}$=0.62
mag. Differential reddening of intra--cluster origin is present.  Pre-main
sequence members are likely to be present over the 6 mag range explored by
our observations (reaching down to earliest A spectral types). The presence
of differential reddening and pre-main sequence members agrees with the
evidence for a bright parent nebulosity embedding the whole cluster. The
radial velocity of the cluster is consistent with the Hron (1987) model of
galactic rotation.

\end{abstract}

\begin{keywords}
Haffner~18: photometry, spectroscopy -- open clusters: HR diagram.
\end{keywords}

\section{Introduction}

Haffner~18 ($\alpha_{2000} = 7^{h}~52_{.}^{m}5$ \ \  $\delta_{2000} =
-26^{o}~22^{\prime}$ \ \ $l~=~243_{.}^{o}1$ \ \ $b~=~+0_{.}^{o}5$) lies
on the sky close to the young open cluster Haffner~19 . Both clusters are
located in a region of Puppis reported by FitzGerald (1968) as remarkably
free of interstellar absorption. They are therefore well suited to trace the
spiral structure and galactic rotation at large distances in the general
anti-center galactic direction. In this paper we present and discuss our
photometric observations of Haffner~18 secured similarly to those on
Haffner~19 (Munari and Carraro 1996a, hereafter MC96a) and another cluster
we have previously studied in the region, Bochum~2 (Munari and Carraro
1995).

Haffner~18 was first identified by Haffner (1957) who sub--divided it into
three distinct groups $a$, $b$ and $c$. FitzGerald and Moffat (1974,
hereafter FM) considered the two clusters Haffner 18a and 18b as a single
elongated entity, parallel to the galactic plane, while the third one was
interpreted as a chance alignment of field stars. FM found the cluster to be
affected by differential reddening and derived a distance modulus of
(m-M)$_\circ=14.2 \pm 0.3$. They reported the HR diagram to show evidence of
pre--main sequence members, arguing for an age $T\leq 10^6$ yrs.  FM also
suggested that Haffner 18 and Haffner~19 are at the same distance and belong
to spiral arm located at 15 Kpc in the anti-galactic center direction,
possibly associated with the Perseus arm extension.

Labhardt et al. (1992, hereafter LSS) presented accurate CCD photometry of
both Haffner~18 and 19. The systematic discrepancy with the FM data was
attributed by them to different treatment of blended companions and reliable
choice of the sky background. Our earlier results on Haffner~19 (MC96a)
supported such a conclusion.

\section {Photometry}

{\sl UBV(RI)$_{\rm C}$} photometry of Haffner~18 has been obtained with the
CCD camera on the 1.0 m telescope of the South Africa Astronomical
Observatory (SAAO) at Sutherland on February 29, 1992. The same night we
observed Haffner~19 on which we already reported elsewhere (MC96a).  The
reader is referred to Munari and Carraro (1995) for the adopted (standard)
procedures and an evaluation of the excellent instrumental performances. 
The journal of observations is given in Table~1 and the resulting {\sl
UBV(RI)$_{\rm C}$} magnitudes are listed in Table~2.  A finding chart is
presented in Figure~1. Star \#1 has been recorded very close to the CCD
frame border, which caused the larger-than-average errors in Table~2.  Star
\#52 appears resolved into two components (LSS~38 and LSS~39) by Labhardt et
al. (1992). Both stars \#1 and 52 will be ignored in the following. On the
other hand, star FM~3067 has been resolved into two objects (\#43 and 44)
and FM~3081 into three (\#55, 56 and 57). Values for star \#53 in Table~2
are nearly identical with those given by LSS (their star 37), whilst FM data
(their star 3073) are largely off. The difference is too large to be
explained in term of observational errors and most probably it is due just
to a printing error. For these reasons stars \#53, FM~3067 and FM~3081 are
not considered in the comparison with FM data in Eq.(1) and (2).

Comparison between our profile CCD photometry (performed with the DAOPHOT package) 
and the FM aperture photoelectric photometry for the 12 stars in common yields:
\begin{eqnarray}
V \ - \ V_{FM}=0.078&& \ \ \ \ \sigma=0.055 \\
(B-V)\ - \ (B-V)_{FM}=0.063 &&\ \ \ \ \sigma=0.052
\end{eqnarray}
\noindent
For the 27 stars in common with the profile CCD photometry by LSS 
(who used the ROMAPHOT package) the comparison gives:
\begin{eqnarray}
V \ - \ V_{LSS}=-0.018 && \ \ \ \ \sigma=0.041 \\
(B-V)\ - \ (B-V)_{LSS}=-0.001 &&\ \ \ \ \sigma=0.051
\end{eqnarray}
\noindent
Details of the comparison are given in graphical form in Figure~2. The
results are fairly similar to those obtained by MC96a for Haffner~19, with
aperture photometry poorly performing in a stellar field as crowded as that
of Haffner~18. On the other hand the comparison with LSS photometry is quite
satisfactory, with the very small residual differences easily accounted for
by the different system of standard stars used (Landolt's equatorial stars
for LSS, Cousins' E-regions for us) and possibly some subtle differences
introduced by the reduction softwares (DAOPHOT vs. ROMAPHOT).

\begin{figure}
\centerline{\psfig{file=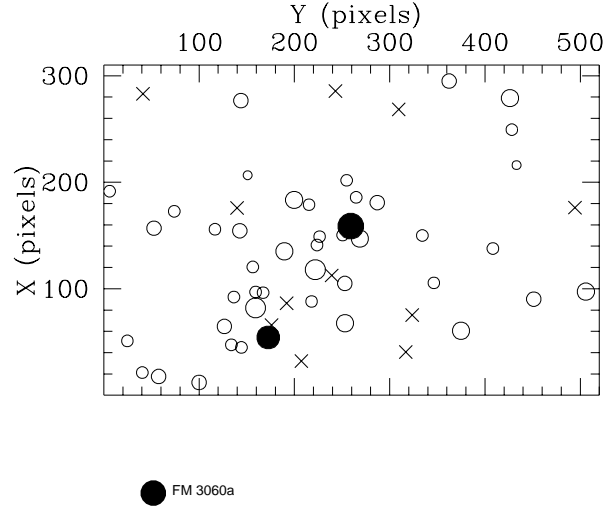,height=8cm,width=8cm}}
\caption[]{Finding chart for the programme stars. The CCD field cover
2.1$\times$3.3~arcmin. North on top, East to the left.
{\sl Crosses}: field stars.
{\sl Open circles}: cluster members. {\sl Filled circles}:
the three earliest type cluster members (O7, O9 and B0.5, the latter
laying outside the region covered by our observations).}
\end{figure}

\begin{figure}
\centerline{\psfig{file=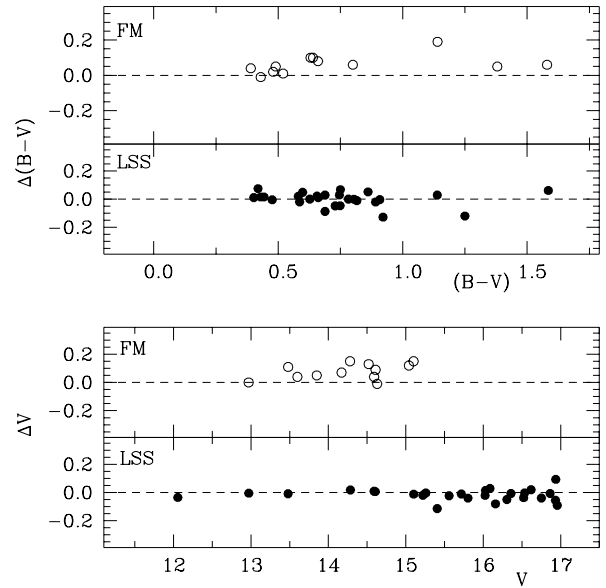,height=8cm,width=8cm}}
\caption[]{Comparison between FM, LSS and our photometry for the stars in 
common.}
\end{figure}

\begin{table}
\tabcolsep 0.10truecm
\caption{Journal of observations.
{\em Seeing} is the FWHM of stellar images as measured on the CCD frames.}
\begin{tabular}{cccc} \hline
\multicolumn{1}{c}{Date} &
\multicolumn{1}{c}{Filter} &
\multicolumn{1}{c}{Exp. time} &
\multicolumn{1}{c}{Seeing} \\
 & &  (sec) & ($\prime\prime$) \\
&&&\\
Feb. 29, 1992 & U &  600 &  2.1\\
& V &   22 &  1.9\\
& V &  600 &  1.8\\
& B &  900 &  1.9\\
& B &   60 &  2.0\\
& R &   10 &  2.1\\
& R &  330 &  1.9\\
& I &   10 &  1.7\\
& I &  600 &  1.9\\
\hline
\end{tabular}
\end{table}

\section{Spectroscopy}

Angela Bragaglia (Astronomical Observatory of Bologna) obtained for us on
Feb. 25, 1994 a long list spectrum of Haffner~18 with the B\&C
+ CCD spectrograph on the ESO 1.5 m telescope in La Silla (Chile). 
The covered range is $\lambda\lambda$ 3380--9190 \AA, the dispersion 
2.5 \AA/pixel and the resolution $\sim$7 \AA. The slit was rotated so 
to record on a single exposure the spectra of stars \#18, 44, 52 and 56 
 plus stars FM~3049, 3059 and 3060a. Emission lines from the underlying 
nebulosity are recorded over the whole slit height ($\sim$4 arcmin ) 
being particularly strong around star FM 3060a. Due to different placement on 
the slit and different brightness, the signal--to--noise ratio of the spectra of the 
individual stars differs greatly, ranging from $\sim$80 to $\sim$11.

Available spectroscopic data on Haffner~18 are summarized in Table~3 (our
and FM classification).  Color excesses and distance moduli have been
computed relative to the intrinsic colors and absolute magnitudes
listed by FitzGerald (1970), Bessell (1990) and Schmidt-Kaler (1982),
assuming the standard Savage \& Mathis (1979) interstellar extinction law.
Empty fields refer to either no available data or cases in which the
interpolation between tabular reference seemed too hazardous. Input
photometry for FM~3049, 3059 and 3060a comes from FM data properly shifted 
to our system.

\begin{table*}
\tabcolsep 0.08truecm
\caption{{\sl UBV(RI)$_{\rm C}$} photometry of Haffner~18. The columns give our
identification number and those used by FM and LSS, the X and Y
positions on the CCD frame, the magnitudes and their associated internal errors
(as provided by the DAOPHOT package; in millimag). For the reasons given in the text, 
stars \#1 and 52 have not been used in the analysis. {\sl Asterisks}: see text.
In the second column a $N$ indicates a star that our analysis suggests to be
a field star.} 
\begin{tabular}{rclcrrrrcrcrcrcr} \hline
ID &&FM &LSS & \multicolumn{1}{c}{X} &  \multicolumn{1}{c}{Y} 
& \multicolumn{1}{c}{V} & $\sigma_V$ & B-V & $\sigma_{B-V}$ & V-I & 
$\sigma_{V-I}$ & V-R & $\sigma_{V-R}$& U-B & $\sigma_{U-B}$  \\
& & & & & & & & & & & & &  \\
   1& &  3090&    & 231.11&   --1.46&   14.672&  25&       &    &  0.432& 25& 0.164& 25&   & \\
   2& &   3089&    & 191.62&     6.16&   16.364&   5&  1.108&  12&  1.495&  6& 0.657& 13&   &\\
   3& &  3053&  25&  51.07&    24.74&   16.158&   4&  0.729&   7&  0.808&  6& 0.382& 14&   &\\
   4& &  3054&  21&  21.34&    40.43&   16.024&   4&  0.661&   6&  0.814&  5& 0.369& 13&   &\\
   5&~N~&3091&    & 283.08&    41.19&   12.430&   0&  0.394&   1&  0.420&  1& 0.201&  1&--0.024&2\\
   6& &  3088&  46& 156.87&    52.66&   15.803&   3&  0.419&   5&  0.444&  6& 0.149& 13&   &\\
   7& &  3055&  20&  17.70&    57.58&   15.258&   2&  0.688&   4&  0.800&  3& 0.377&  7&   &\\
   8& &      &  47& 172.75&    73.98&   16.751&  13&  0.816&  18&  0.927& 15&      &   &   &\\
   9& &  3056&  19&  12.09&   100.01&   15.102&   2&  0.656&   4&  0.796&  3& 0.379&  5&   &\\
  10& &      &  45& 155.93&   116.66&   16.956&  14&  0.908&  21&  0.975& 17&      &   &   &\\
  11& &  3069&  29&  64.71&   126.52&   15.559&   3&  0.443&   4&  0.610&  5& 0.263&  9&   &\\
  12& &      &  34&  92.29&   136.54&   16.933&   8&  0.598&  13&  0.804& 13&      &   &   &\\
  13&N&  3077&  48& 175.85&   139.75&   12.972&   1&  1.585&   2&  1.641&  1& 0.811&  1&   &\\
  14& &  3076&  44& 154.47&   142.70&   15.717&   3&  0.861&   5&  1.171&  6& 0.560&  9&   &\\
  15& &      &    & 206.65&   150.97&   17.162&   9&  0.658&  17&  0.932& 13&      &   &   &\\
  16& &  3071&  40& 120.49&   156.35&   16.358&   5&  1.250&  13&  1.547&  6& 0.743& 12&   &\\
  17& &  3075&  41& 135.13&   189.49&   14.589&   1&  0.627&   2&  0.973&  1& 0.412&  3&   &\\
  18&N& 3072a&  32&  86.51&   191.84&   14.284&   3&  0.804&   4&  1.064&  4& 0.496&  4&   &\\
  19&N&  3066&  22&  32.18&   207.28&   14.607&   1&  1.139&   3&  1.256&  1& 0.599&  3&   &\\
  20& &      &  33&  88.10&   218.05&   16.534&   6&  0.921&  11&  1.307&  7&      &   &   &\\
  21&N&  3085&    & 285.62&   243.16&   11.839&   1&  1.084&   1&  0.999&  1& 0.531&  1&  0.607&4\\
  22& &  3065&    &  67.69&   253.10&   14.631&   1&  0.640&   2&  0.665&  1& 0.331&  3&   &\\
  23& &  3080&    & 201.84&   254.84&   16.464&   5&  0.530&   9&  0.737&  9&      &   &   &\\
  24& &      &    & 185.91&   264.80&   16.961&  14&  0.760&  25&  1.097& 22&      &   &   &\\
  25& &  3082&    & 180.93&   286.99&   15.041&   4&  0.489&   4&  0.725&  5& 0.306&  6&   &\\
  26&N&  3084&    & 268.37&   309.48&    9.901&   4&  1.119&   6&  1.084&  8& 0.512&  6&  0.790&5\\
  27&N&  3064&    &  40.56&   316.85&   13.850&   1&  0.481&   1&  0.563&  1& 0.261&  2&  0.063&10\\
  28&N&  3063&    &  75.31&   323.64&   13.598&   1&  1.377&   2&  1.424&  1& 0.707&  2&   &\\
  29& &      &    & 150.04&   334.21&   17.000&   9&  0.872&  15&  0.937& 11&      &   &   &\\
  30& &      &    & 105.55&   346.37&   16.923&   9&  0.651&  14&  0.734& 11&      &   &   &\\
  31& &  3083&    & 295.09&   362.36&   15.659&   3&  0.668&   5&  0.797&  4& 0.357&  9&   &\\
  32& &  3062&    &  60.61&   374.86&   14.169&   2&  0.522&   2&  0.608&  3& 0.290&  4&   &\\
  33& &      &    & 137.79&   408.30&   16.931&  10&  0.739&  16&  0.962& 21&      &   &   &\\
  34& &  2057&    & 279.02&   426.23&   14.520&   1&  0.583&   2&  0.651&  2& 0.302&  3&   &\\
  35& &      &    & 249.43&   428.16&   16.458&   5&  1.029&  10&  1.410&  5& 0.690& 14&   &\\
  36& &      &    & 216.13&   433.09&   17.118&   9&  0.631&  16&  0.742& 13&      &   &   &\\
  37& &      &    &  90.30&   451.13&   15.925&   3&  0.632&   6&  0.758&  4& 0.350& 12&   &\\
  38&N&      &    & 176.22&   494.45&   13.720&   1&  0.594&   1&  0.706&  1& 0.335&  2&  0.112&10\\
  39& &      &    &  97.33&   505.83&   14.594&   1&  0.589&   2&  0.685&  2& 0.406& 59&   &\\
  40& &  3068&  24&  47.38&   133.86&   16.305&  12&  0.581&  19&  0.798& 18&      &   &   &\\
  41& &      &  23&  44.94&   144.39&   16.522&  18&  0.586&  25&  0.931& 27&      &   &   &\\
  42& &  3086&    & 276.67&   143.90&   15.701&   3&  0.387&   4&  0.413&  5&      &   &   &\\
  43& & 3067*&  27&  54.34&   172.59&   12.055&   1&  0.402&   2&  0.598&  3& 0.263&  1&--0.559&2\\
  44&N& 3067*&  28&  62.83&   175.97&   15.406&   6&  0.688&  23&       &   & 0.072& 18&   &\\
  45& &  3078&    & 183.46&   199.76&   14.517&   2&  0.391&   3&  0.584&  2& 0.241&  4&--0.380&2\\
  46& &  3079&  49& 178.95&   215.35&   16.029&   7&  0.476&   9&  0.546&  9& 0.229& 16&   &\\
  47& &      &  42& 141.11&   223.74&   16.618&   7&  0.782&  12&  0.890&  9&      &   &   &\\
  48& &      &  43& 148.99&   226.47&   16.864&   9&  0.749&  17&  1.177& 11&      &   &   &\\
  49& &  3070&  31&  82.09&   159.24&   13.479&   1&  0.429&   1&  0.603&  1& 0.260&  2&--0.443&4\\
  50& &      &  36&  96.98&   159.33&   16.087&   6&  0.746&   9&  0.844&  8& 0.417& 25&   &\\
  51& &      &  35&  96.38&   167.13&   16.936&  14&  0.750&  24&  0.906& 19&      &   &   &\\
  52& & 3074&38,39*&117.96&   221.94&   13.025&   1&  0.158&   3&  0.250&  5& 0.093&  2&--0.236&5\\
  53&N&  3073&  37& 112.51&   239.13&   15.220&   3&  0.891&  11&  1.096&  4& 0.590& 12&   &\\
  54& & 3072b&    & 105.03&   252.90&   15.624&   3&  0.468&   5&  0.833&  4& 0.465&  9&   &\\
  55& & 3081*&    & 150.43&   250.66&   16.153&  20&  0.546&  33&  0.991& 33&      &   &   &\\
  56& & 3081*&    & 158.83&   259.14&   11.212&   1&  0.333&   3&  0.453&  2& 0.183&  1& --0.634&3\\
  57& & 3081*&    & 147.11&   268.75&   14.933&   8&  0.769&  14&  0.967&  9& 0.527& 12&   &\\
\hline           
\end{tabular}    
\end{table*}     

\section{Results}

\subsection{Color-Magnitude diagram}

The {\sl V-(B-V)} and {\sl V-(V-I$_C$)} diagrams are shown in Figure~3. The
main sequence appears well defined down to the limit of our photometry in
Table~2 ($V$ = 17.2 mag). Star \#26 (a G8~V) is clearly a foreground object (cf.
Table~3) and lies outside the diagram limits in Figure~3.  Labeled stars
(\#13, 19, 21 and 28) are classified as field stars from their
location in the color-magnitude diagrams. It has also to be noted that
according to the preliminary spectroscopic results of Table~3 stars \#18 
and 44 are not members if their luminosity class is taken to be V.

Our {\sl V-(B-V)} and {\sl V-(V-I$_C$)} diagrams are affected by errors not
exceeding $\sigma$=0.015 mag over the explored range, and therefore the
widening of the main sequence with fainter magnitudes is a real effect that
we ascribed to the combined effects of differential reddening and presence
of pre--main sequence stars. Similar conclusions were reached by FM. A few
field stars could be still hidden in the cluster main sequences of Figure~3.
We secured photometry for a couple of field regions around Haffner~18 to the
aim of statistically clean the diagrams in Figure~3. However, the strong
intra--cluster extinction, reddening and emission nebulosity prevent our attempt 
to perform the exercise in a straightforward manner.

\begin{figure}
\centerline{\psfig{file=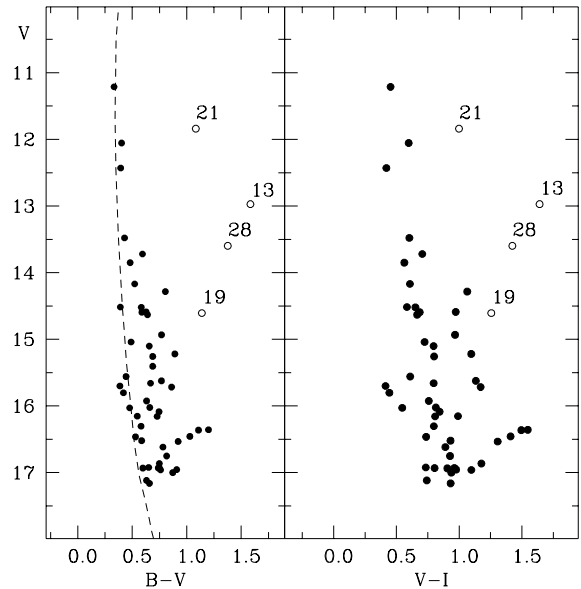,height=8cm,width=8cm}}
\caption[]{{\sl V-(B-V)} and {\sl V-(V-I$_C$)} color magnitude diagrams
from Table~2. Probable field stars are labeled. The dotted line is 
a solar abundance $T=2 \times 10^{6}$ yr isochrone from the Padova group
(Bertelli et al. 1994).}
\end{figure}

\subsection{Distance, Reddening and Age}

Distance modulus and mean reddening have been obtained through trial fits to the
observed color-magnitude diagram with theoretical isochrones of the Padova
group ({\em cf.} Bertelli {\em et al.} 1994) characterized by a standard
$[$He/H$]$ ratio and $Z$=0.020.

To better estimate distance and  color excess of Haffner~18 we fitted in
Figure~3 the locus  of main sequence stars with minimal reddening. The fit
is good over the whole blue side of the main sequence,
particularly for the O stars on the upper sequence. Their stronger winds
could have already blown away from their surroundings the residuals of the
parent cloud, thus reducing towards pure interstellar terms the reddening
affecting them. 

The best fit is achieved for an age  $T = 2\times 10^6$ yrs, a reddening
$E_{B-V}=0.62 \pm 0.08$  and a distance modulus (m-M)$_\circ =14.0 \pm 0.2$
mag (corresponding to a distance of 6.3 kpc). This is fairly compatible with
the average spectroscopic distance modulus of the four OB stars in Table~3
which is (m-M)$_\circ = 14.57 \pm 0.94$ mag. FM estimated from fit to their
ZAMS a value (m-M)$_\circ = 14.2 \pm 0.3$ mag. The three determinations well
agree inside the errors. Comparison with the results by MC96a suggests that
Haffner~18 and Haffner~19 are located at significantly different distances,
appearing close on the sky just by chance alignment. FM on the contrary
argued for Haffner~18 and 19 to lay close in space.

The $Q-$method (Becker 1963) has been applied to the stars in Table~2 with
$U$ band photometry available. The resulting reddening free color-magnitude
diagram is presented in Figure~4. Star \#53 and FM 3060a have been also
plotted, their de--reddened values inferred from the spectroscopy in Table~3.
The excellent fit to a $2\times 10^6$ yrs isochrone scaled to
(m-M)$_\circ=14.0$ mag is evident as well the segregation of several field
stars. The location of star FM 3060a cannot be reconciled
with the B1~V classification by FM, its position being in excellent
agreement with the B0.5~III classification given in Table~3.

The mean reddening for the cluster members in Figure~4 (the O-B earliest
members) is $E_{B-V} = 0.66 \pm 0.03$. Another way to estimate the {\sl
mean} reddening over the cluster (thus including the differential reddening)
is through a linear fit to the main sequence in the {\sl B--V, B--I$_C$}
plane as suggested by Natali et al. (1994) and revised by Munari and Carraro
(1996b). Since this method applies only to main sequence stars, we tried to
avoid pre main sequence objects by using only those stars that in Figure~3
lie closer than $\bigtriangleup B-V$=0.15 to the fitting isochrone. The
linear fit on the {\sl B--V, B--I$_C$} plane (see Figure~5) as described by
Munari and Carraro (1996b) for the R$_V$=3.1 extinction law
\begin{equation}
(B-I) \ = \ \Omega \ + \ 2.25 (B-V)
\end{equation}
leads to the following expression for the color excess:
\begin{equation}
E_{B-V} = \frac{\Omega - 0.014}{0.159}
\end{equation}
The fit in Figure~5 gives $\Omega$=0.129 which implies E$_{B-V}$=0.72, close to
the mean value over the cluster E$_{B-V}$=0.70 reported by FM and in fair
agreement with the above derived values ( 0.62 and 0.66).

\subsection{Membership}

We have surveyed 57 stars in a region $2.1^\prime \times 3.3^\prime$ roughly
centered on the cluster core. Excluding stars \#1 and \#52 for the reasons
given above, among the remaining 55 stars we have above identified 11
non-members (marked with an $N$ in the second column of Table~2). The
remaining 44 stars are taken as cluster members once differential reddening
and pre--main sequence objects are taken into account.

\begin{figure}
\centerline{\psfig{file=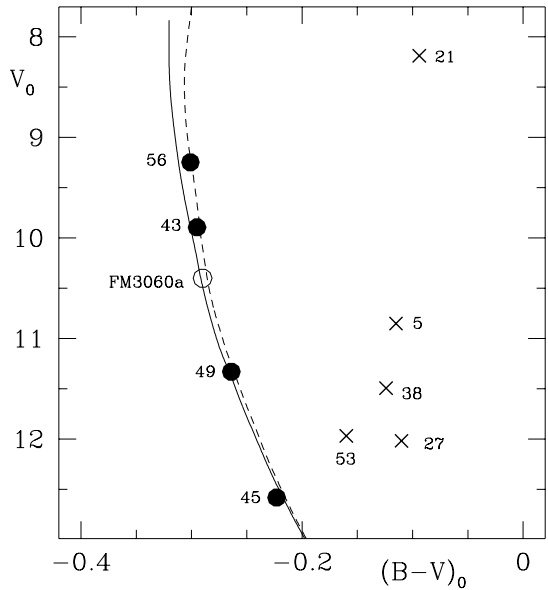,height=8cm,width=8cm}}
\caption[]{Location on the {\sl V$_\circ$-(B-V)$_\circ$} diagram of the
programme stars with available $U$ band photometry de reddened via the
Q--method. Field stars are indicated with crosses. 
The solid and dashed lines are the ZAMS and a solar abundance
2$\times 10^6$ yr isochrone from the Padova group (Bertelli et al. 1994), 
scaled to (m-M)$_\circ$ = 14.0 mag.}
\end{figure}

\begin{figure}
\centerline{\psfig{file=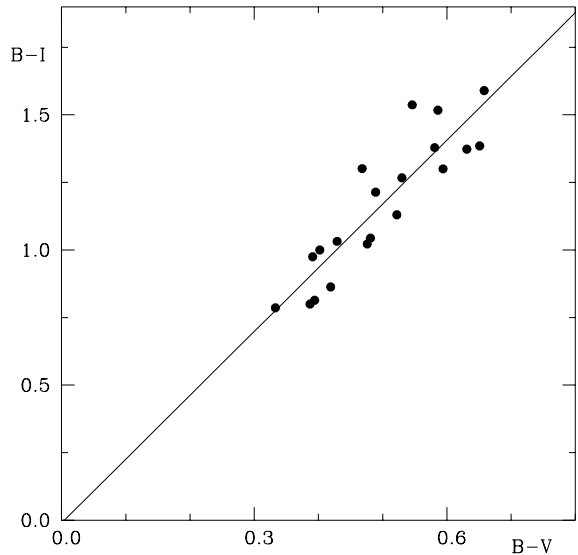,height=8cm,width=8cm}}
\caption[]{Haffner~18 members plotted on the (B-V)-(B-I$_C$) plane.
The solid line is the least square fit described in sect. 4.2}
\end{figure}

\begin{table}
\tabcolsep 0.10truecm
\caption{Spectroscopic classification, reddening and spectroscopic
parallaxes. {\sl FM}: numbering and classification from Fitzgerald
and Moffat (1974). See sect.~2 for a warning on star \#52.}
\begin{tabular}{llllccccr} 
\hline
\multicolumn{2}{c}{this paper}&\multicolumn{2}{c}{FM}&&&&&\\
&&&&$E_{B-V}$&$E_{U-B}$&$E_{V-R}$&$E_{V-I}$&(m-M)$_\circ$\\
   &          &        &       &      &      &      &      &      \\
 18&A4        &  3072a &       & 0.68 &      & 0.45 & 0.95 & 10.4 \\
 26&          &  3084  & G8 V  & 0.38 & 0.49 & 0.08 & 0.27 &  3.2 \\
 43&          &  3067  & O9 IV & 0.71 & 0.57 & 0.42 & 0.93 & 15.1 \\
 44&A5--6     &        &       & 0.53 &      & 0.01 &      & 11.6 \\
 52&A0 II/III &  3074  &       &      &      &      &      &      \\
 53&          &  3073  & B5:   & 1.05 &      & 0.62 & 1.19 & 13.2 \\
 56&O6        &  3081  & O7:   & 0.65 & 0.55 & 0.34 & 0.78 & 14.7 \\
   &F0 II:    &  3049  &       & 0.25 & 0.0  &      &      & 17.9 \\
   &A2        &  3059  &       & 0.14 & 0.03 &      &      & 12.8\\
   &B0.5 III  &  3060a & B1 V  & 0.50 & 0.33 &      &      & 15.3 \\
\hline
\end{tabular}
\end{table}

\subsection{Galactic rotation}

FM have reported preliminary radial velocities for three stars of
Haffner~18. Dropping star FM~3081 which is probably a binary (FM
suspected the spectrum to be doubled--lined), the remaining two stars
give a heliocentric radial velocity of $\sim 67\pm 10$ km sec$^{-1}$ for
Haffner~18. FM reported that this velocity gives a kinematic distance of 4.5
kpc based on the Schmidt (1965) rotation model of the Galaxy, which is
significantly shorter than 6.9 kpc distance they derived from fitting to 
the ZAMS.

The heliocentric component of the cluster radial velocity due to the 
galactic rotation can be expressed as:
\begin{eqnarray}
RV_\odot & = & w_\circ\ {\rm sin}b \ + \ u_\circ\ {\rm cos}l\ {\rm cos}b 
               \ + \ v_\circ\ {\rm sin}l\ {\rm cos}b \nonumber \\
         &   & - \ 2[A(R - R_\circ) + \alpha (R-R_\circ)^2]\ {\rm sin}l\ {\rm cos}b 
\end{eqnarray}
where as usual $A$ denotes the Oort's constant, $\alpha$ is the curvature
term (cf. Hron 1987), R,R$_\circ$ are the cluster and Sun galactocentric
distances, {\sl d} the cluster--Sun distance, ({\em l,b}) the heliocentric
galactic coordinates of the cluster and ($u_\circ, v_\circ, w_\circ$) is the
solar motion vector. Adopting our estimate of 6.3 kpc for the distance and
($u_\circ, v_\circ, w_\circ$)=(9.3,11.2,7.0) km sec$^{-1}$ from Pont et al.
(1994), Eq.(7) rewrites for Haffner~18 as:
\begin{eqnarray}
RV_\odot & = & -14.1 \ +\ 7.4 A\ +\ 31.0 \alpha  
\end{eqnarray}
Hron (1987) has used distances and radial velocities to young open clusters
to investigate the rotation curve of the Galaxy. His results valid for the
range $-3 < R - R_\circ < 5$ kpc are best fitted by A=$17.0 \pm 1.5$ km
sec$^{-1}$ kpc$^{-1}$ and $\alpha = -2.0 \pm 0.6$ km sec$^{-1}$ kpc$^{-2}$. 
If we insert them in Eq.(8) the radial velocity of Haffner~18 results to be
50 km sec$^{-1}$, which marginally agrees inside the errors with the $\sim 67\pm
10$ km sec$^{-1}$ given by FM. To obtain exactly 67 km sec$^{-1}$ from the Hron's
rotation curve it is enough to lower the curvature term to $\alpha = -1.44$,
a value well compatible with the uncertainty given by Hron.

In this sector of the galactic plane the radial component of the galactic
motion is however not very sensitive to the Sun--cluster distance. 
We may nevertheless conclude that there is no discrepancy between the photometric,
spectroscopic and kinematic independent determinations of the distance to
Haffner~18.

\section*{Acknowledgments}
One of us (UM) was Visiting Astronomer at the South African Astronomical
Observatory (Cape Town) when the observations described in this paper were
collected. This work has been financially supported by the Italian Ministry
of University, Scientific Research and Technology (MURST) and the Italian
Space Agency (ASI).


\begin{thebibliography}{}
\bibitem{} Bertelli G., Bressan A., Chiosi C., Fagotto F., Nasi E., 1994,
           A\&AS 106, 275
\bibitem{} Becker W. 1963, in {Stars and Stellar Systems: Basic Astronomical
           Data}, K.A.Strand ed., Univ. of Chicago Press, pag. 241
\bibitem{} Bessell M. S., 1990, PASP 102, 1181
\bibitem{} FitzGerald M. P., 1968, A. J. 73, 983
\bibitem{} FitzGerald M. P., 1970 A\&A 4, 234
\bibitem{} FitzGerald M. P., Moffat A. F. J., 1974, AJ 79, 873 (FM)
\bibitem{} Haffner H., 1957, Z.f.Astrophysik 43, 89
\bibitem{} Hron J., 1987, A\&A 176, 34
\bibitem{} Labhardt L., Spaenhauer A., Schwengeler H., 1992, A\&A 265, 869  (LSS)
\bibitem{} Munari U., Carraro G., 1995, MNRAS 277, 1269
\bibitem{} Munari U., Carraro G., 1996a, MNRAS 283, 905  (MC96a) 
\bibitem{} Munari U., Carraro G., 1996b, A\&A 314, 108
\bibitem{} Natali F., Natali G., Pompei E., Pedichini F., 1994, A\&A 289, 756
\bibitem{} Pont F., Mayor M, Burki G., 1994, A\&A 285, 415
\bibitem{} Savage B.D., Mathis J.S. 1979, ARA\&A 17, 73
\bibitem{} Schmidt M. 1965, in {\sl Stars and Stellar Systems: 
           Galactic Structure}, ed. A. Blaauw and 
           M. Schmidt, Univ. of Chicago Press, p. 513
\bibitem{} Schmidt-Kaler, Th. 1982 in Landolt-B\"{o}rnstein, Numerical Data
            and Functional Relationships in Science and Technology, ed. K.Schaifer 
            \& H.H. Voigt, New Series, Group IV, Vol.II(b) (Pringer, Berlin), pag. 14
\end{thebibliography}
\end{document}